\documentclass[aps,preprint,nofootinbib]{revtex4}
\usepackage{amsmath}
\usepackage{graphicx}
\usepackage{color}
\usepackage{slashed}
\usepackage[aboveskip=-5pt]{subcaption}
\usepackage{booktabs}
\usepackage{mathtools}
\usepackage{multirow}
\usepackage{makecell}
\allowdisplaybreaks

\begin{document}
\title{Exclusive $J/\psi+\gamma$ production in ultraperipheral ion collisions}

\author{\vspace{1cm} Zi-Qiang Chen\footnote[1]{chenziqiang13@gzhu.edu.cn} and Long-Bin Chen\footnote[2]{chenlb@gzhu.edu.cn, corresponding author}}

\affiliation{School of Physics and Materials Science, Guangzhou University, Guangzhou 510006, China}

\begin{abstract}
\vspace{0.5cm}
Ultraperipheral collisions (UPCs) of ions provide new opportunities to study the quarkonium production mechanism in photon-photon scattering.
In this paper, we investigate the exclusive process $\gamma+\gamma\to J/\psi+\gamma$ up to $\mathcal{O}(\alpha_s v^2)$ accuracy within the nonrelativistic quantum chromodynamics factorization framework.
We evaluate the corresponding cross sections for Pb-Pb and p-p UPCs at the Large Hadron Collider.
Numerical results show that the $\mathcal{O}(\alpha_s)$, $\mathcal{O}(v^2)$, and $\mathcal{O}(\alpha_s v^2)$ corrections are about $-50\%$, $-33\%$, and $15\%$ of the leading-order (LO) contribution, respectively, showing reasonable convergence  in both $\alpha_s$ and $v^2$ expansion.
Collectively, these corrections suppress the LO cross section by a factor of about $1/3$, which is a crucial effect for reliable phenomenological analysis. 
Our results suggest that future experimental measurements of this process are feasible.
\end{abstract}
\maketitle

\newpage

\section{Introduction}
Heavy quarkonium plays an important role in high energy collider physics, as it provides an ideal laboratory for exploring the interplay between perturbative and nonperturbative quantum chromodynamics (QCD). 
Owing to the large mass of its constituent quarks, the quarkonium production process is assumed to be factorizable into two stages. 
First, a heavy quark-antiquark pair with an invariant mass near the bound-state mass is produced perturbatively, then the pair binds nonperturbatively into the quarkonium state.
This process can be systematically described by the non-relativistic QCD (NRQCD) factorization formalism \cite{Bodwin:1994jh}, where the nonperturbative hadronization effects are encoded into the long distance matrix elements (LDMEs).
Theoretical predictions within the NRQCD factorization framework takes the form of a double expansion in the strong coupling constant $\alpha_s$ and the heavy quark relative velocity $v$.

While rooted in rigorous effective field methods, the NRQCD factorization formalism still requires to be verified in various collider experiments.
Quarkonium production in photon-photon collisions provides an interesting testing ground. 
The cross section for inclusive $J/\psi$ production via photon-photon scattering was measured by the LEP II experiment \cite{DELPHI:2003hen}, yet the underlying production mechanism remains incompletely understood. 
Theoretical studies indicate that the color-singlet (CS) contributions alone are inadequate to explain the data, while the predictions to color-octet (CO) contributions based on different LDME sets diverse quite a lot, which even lead to opposite conclusions in some cases \cite{Klasen:2001cu,Butenschoen:2011yh,Brambilla:2024iqg}.
Investigating specific subprocesses separately may help to clarify the situation \cite{Ma:1997bi,Klasen:2001mi,Qiao:2003ba,Klasen:2004tz,Klasen:2004az,Li:2009zzu,Chen:2016hju}. while experimental data for such channels is currently limited by the scarcity of the events at existing $e^+e^-$ collider.
Fortunately, ultraperipheral collisions (UPCs) at the Large Hadron Collider (LHC) provide new opportunities for the study of photon-photon physics.
In UPC, the ion impact parameter is larger than twice the ion radius, hence the ions are kept unbroken during the collision, and the interaction is pure electromagnetic.
This allows photons from the colliding ions to interact at high center-of-mass energy.
Furthermore, with the luminosity upgrade of the LHC, both inclusive and exclusive processes can be measured with high precision.

In this paper, we focus on the exclusive process $\gamma+\gamma\to J/\psi +\gamma$.
Compared to inclusive $J/\psi$ production, the production mechanism of $J/\psi+\gamma$ is relatively simpler, which may provide key insight into the quarkonium production process in photon-photon scattering.
This process was previously investigated at the next-to-leading-order (NLO) in $\alpha_s$, as a subprocess of $\gamma+\gamma\to J/\psi +\gamma+X$, in Ref. \cite{Klasen:2004az}.
While the phenomenological analysis assumed a future $e^+e^-$ linear collider, which is not yet realized.
Recently, the authors of Ref. \cite{Goncalves:2023sts} studied this process in Pb-Pb UPC, at the leading-order (LO) accuracy.
Their LO results confirm the feasibility of measuring this process at the High-Luminosity LHC (HL-LHC).
Considering that the QCD and relativistic corrections to charmonium production process are normally significant, to make a more reliable prediction, in this work, we perform a comprehensive calculation by including the $\mathcal{O}(\alpha_s)$, $\mathcal{O}(v^2)$, and  $\mathcal{O}(\alpha_s v^2)$ corrections to the $\gamma+\gamma\to J/\psi +\gamma$ process.

The rest of the paper is organized as follows. 
In Section II, we present the general formula for $\gamma+\gamma\to J/\psi+\gamma$ process up to $\mathcal{O}(v^2)$ within the NRQCD factorization framework.
In Section III, we elucidate some technical details for the perturbative calculation. 
In Section IIII, the numerical evaluation for concerned process is performed. The last section is a summary and conclusions.

\section{NRQCD factorization formula up to $\mathcal{O}(v^2)$}
According to the NRQCD factorization formula \cite{Bodwin:1994jh,Bodwin:2010fi}, the amplitude for $\gamma+\gamma\to J/\psi+\gamma$ process up to $\mathcal{O}(v^2)$ can be expressed as 
\begin{equation}
\mathcal{M}^{\gamma+\gamma\to J/\psi+\gamma}=\sqrt{2M}\;\left[c_0\; \langle J/\psi| \psi^\dagger \boldsymbol{\sigma}\cdot \boldsymbol{\epsilon} \chi |0\rangle+\frac{c_2}{m^2}\; \langle J/\psi | \psi^\dagger (-\tfrac{i}{2}\overleftrightarrow{\boldsymbol{D}})^2\boldsymbol{\sigma}\cdot \boldsymbol{\epsilon} \chi|0\rangle\right],
\label{eq_amp1}
\end{equation}
where, $c_0$ and $c_2$ are short-distance coefficients, $\langle J/\psi | \psi^\dagger\boldsymbol{\sigma}\cdot \boldsymbol{\epsilon} \chi |0\rangle$ and $\langle J/\psi | \psi^\dagger (-\tfrac{i}{2}\overleftrightarrow{\boldsymbol{D}})^2\boldsymbol{\sigma}\cdot \boldsymbol{\epsilon} \chi|0\rangle$ are NRQCD vacuum-to-$J/\psi$ matrix elements.
The quantities $m$ and $M$ represent the masses of charm quark and $J/\psi$ meson, respectively.
The prefactor $\sqrt{2M}$ on the right side of Eq. (\ref{eq_amp1}) reconciles the non-relativistic normalization convention of NRQCD matrix elements with the relativistic normalization of the amplitude on the left side. 
Note, both $\langle J/\psi | \psi^\dagger\boldsymbol{\sigma}\cdot \boldsymbol{\epsilon} \chi |0\rangle$ and $\langle J/\psi | \psi^\dagger (-\tfrac{i}{2}\overleftrightarrow{\boldsymbol{D}})^2\boldsymbol{\sigma}\cdot \boldsymbol{\epsilon} \chi|0\rangle$ are independent of $J/\psi$ helicity, and there are no helicity summation in Eq. (\ref{eq_amp1}).

To determine the short-distance coefficients $c_0$ and $c_2$ in Eq. (\ref{eq_amp1}),  we follow the standard matching procedure \cite{Bodwin:1994jh}. 
Since these coefficients are insensitive to long-distance dynamics, we replace the physical $J/\psi$ state with a free $c\bar{c}$ pair carrying the quantum number $^3S_1^{[1]}$.
This allows us to exploit the matching condition between perturbative QCD and perturbative NRQCD:
\begin{align}
\mathcal{M}^{\gamma+\gamma\to c\bar{c}(^3S_1^{[1]})+\gamma}\Big|_\text{pert QCD}=&c_0\;\langle c\bar{c}(^3S_1^{[1]})|\psi^\dagger\boldsymbol{\sigma}\cdot \boldsymbol{\epsilon}\chi |0\rangle\Big|_\text{pert NRQCD}\nonumber \\
&+\frac{c_2}{m^2}\; \langle c\bar{c}(^3S_1^{[1]}) | \psi^\dagger (-\tfrac{i}{2}\overleftrightarrow{\boldsymbol{D}})^2\boldsymbol{\sigma}\cdot \boldsymbol{\epsilon} \chi|0\rangle\Big|_\text{pert NRQCD}.
\label{eq_match}
\end{align}
Here, the factor $\sqrt{2M}$ is absent because we use the same (relativistic) normalization for the $c\bar{c}$ state on both sides.
The calculation details of $\mathcal{M}^{\gamma+\gamma\to c\bar{c}(^3S_1^{[1]})+\gamma}\Big|_\text{pert QCD}$ will be presented in Section III.
For now, we express it formally as an expansion series
\begin{equation}
\mathcal{M}^{\gamma+\gamma\to c\bar{c}(^3S_1^{[1]})+\gamma}\Big|_\text{pert QCD}=\mathcal{M}_{00}+\frac{\alpha_s}{\pi}\mathcal{M}_{01}+\frac{ \boldsymbol{q}^2}{m^2}\mathcal{M}_{10}+\frac{ \boldsymbol{q}^2}{m^2}\frac{\alpha_s}{\pi}\mathcal{M}_{11},
\label{eq_mexp}
\end{equation}
where $\boldsymbol{q}$ is the three momentum of the $c$ quark in the $c\bar{c}$ rest frame, and therefore the velocity variable $v=|\boldsymbol{q}|/m$.
It should be note that in calculating $\mathcal{M}_{01}$ and $\mathcal{M}_{11}$, we expand the amplitude as series in $\boldsymbol{q}^2$, before performing the Feynman integration.
This procedure avoids encountering the Coulomb singularity, which would manifest as $\mathcal{O}(1/v)$ terms if the $\boldsymbol{q}^2$ expansion were performed after the Feynman integration.
In dimensional regularization with spacetime dimension $D=4-2\epsilon$, the NRQCD vacuum-to-$c\bar{c}$ matrix elements up to $\mathcal{O}(\alpha_s v^2)$ take the form \cite{Bodwin:1994jh,Jia:2011ah,Guo:2011tz,Li:2013otv}
\begin{align}
\label{eq_pnr1}
&\langle c\bar{c}(^3S_1^{[1]})|\psi^\dagger\boldsymbol{\sigma}\cdot \boldsymbol{\epsilon} \chi |0\rangle\Big|_\text{pert NRQCD}=\sqrt{2N_c}2E_{\boldsymbol{q}}\bigg[1+\frac{2C_F}{3}\frac{\boldsymbol{q}^2}{m^2}\frac{\alpha_s}{\pi} \left(\frac{\mu_\text{R}^2}{\mu_\text{F}^2}\right)^\epsilon\nonumber \\
& \quad\quad\quad\quad\quad\quad\quad\quad\quad\quad\quad\quad\quad\quad\quad\quad\times\left(\frac{1}{\epsilon_\text{IR}}-\gamma_\text{E}+\ln (4\pi)\right) \bigg],\\
\label{eq_pnr2}
&\langle c\bar{c}(^3S_1^{[1]}) | \psi^\dagger (-\tfrac{i}{2}\overleftrightarrow{\boldsymbol{D}})^2\boldsymbol{\sigma}\cdot \boldsymbol{\epsilon} \chi|0\rangle\Big|_\text{pert NRQCD}=\sqrt{2N_c}2E_{\boldsymbol{q}}\boldsymbol{q}^2,
\end{align}
where $N_c=3$, $C_F=4/3$ are color factors, $E_{\boldsymbol{q}}=\sqrt{m^2+\boldsymbol{q}^2}$ is the energy of the $c$  (or $\bar{c}$) quark in the $c\bar{c}$ rest frame, $\mu_\text{F}$ and $\mu_\text{R}$ denote the factorization and renormalization scales, respectively.
Note that the NRQCD operator in Eq. (\ref{eq_pnr1}) is renormalized in the modified minimal-subtraction ($\overline{\text{MS}}$) scheme.
Substituting Eqs. (\ref{eq_mexp})-(\ref{eq_pnr2}) into Eq. (\ref{eq_match}), the short-distance coefficients $c_0$ and $c_2$ can be solved order by order in $v^2$ and $\alpha_s$ expansion.

With the amplitude in Eq. (\ref{eq_amp1}), we obtain the cross section for $\gamma+\gamma\to J/\psi+\gamma$ process
\begin{align}
\hat{\sigma}_\text{NLO}^{\gamma+\gamma\to J/\psi+\gamma}&=\frac{1}{2\hat{s}}\int \text{d} \text{PS}_2\; \overline{\sum} \left|\mathcal{M}^{\gamma+\gamma\to J/\psi+\gamma} \right|^2 \nonumber \\
&=\frac{1}{2\hat{s}}\int \text{d} \text{PS}_2\;\frac{\langle\mathcal{O}_1 \rangle_{J/\psi}}{6m}\left(\mathcal{S}_{00}+\frac{\alpha_s}{\pi}\mathcal{S}_{01}+\frac{\langle\boldsymbol{q}^2\rangle}{m^2}\mathcal{S}_{10}+\frac{\langle\boldsymbol{q}^2\rangle}{m^2}\frac{\alpha_s}{\pi}\mathcal{S}_{11}\right),
\label{eq_crss}
\end{align}
where $\hat{s}$ is the center-of-mass energy squared of the colliding photons, $\text{d}\text{PS}_2$ denotes the two-body phase space measure, and $\overline{\sum}$ indicates summation over final-state spins and averaging over initial-state spins.
The LDME $\langle\mathcal{O}_1 \rangle_{J/\psi}=\left|\langle J/\psi| \psi^\dagger \boldsymbol{\sigma}\cdot \boldsymbol{\epsilon} \chi |0\rangle\right|^2$ and the ratio $\langle \boldsymbol{q}^2\rangle=\frac{\langle J/\psi | \psi^\dagger (-\frac{i}{2}\overleftrightarrow{\boldsymbol{D}})^2\boldsymbol{\sigma}\cdot \boldsymbol{\epsilon} \chi|0\rangle}{\langle J/\psi | \psi^\dagger\boldsymbol{\sigma}\cdot \boldsymbol{\epsilon} \chi |0\rangle}$ are taken as input parameters.
After the matching procedure, we obtain
\begin{align}
\mathcal{S}_{00}&=\overline{\sum}\left|\mathcal{M}_{00}\right|^2, \\
\mathcal{S}_{01}&=\overline{\sum}2\text{Re}\left(\mathcal{M}_{01}\mathcal{M}_{00}^*\right),\\
\mathcal{S}_{10}&=\overline{\sum}2\text{Re}\left(\mathcal{M}_{10}\mathcal{M}_{00}^*\right)-\frac{\mathcal{S}_{00}}{2}, \\
\mathcal{S}_{11}&=\overline{\sum}2\text{Re}\left(\mathcal{M}_{10}\mathcal{M}_{01}^*+\mathcal{M}_{11}\mathcal{M}_{00}^* \right)-\frac{4C_F}{3}\left(\frac{\mu_\text{R}^2}{\mu_\text{F}^2}\right)^\epsilon \left(\frac{1}{\epsilon_\text{IR}}-\gamma_\text{E}+\ln (4\pi)\right)\mathcal{S}_{00}-\frac{\mathcal{S}_{01}}{2}.
\label{eq_sij}
\end{align}
Throughout this paper, we designate the contributions from $\mathcal{S}_{00}$, $\mathcal{S}_{01}$, $\mathcal{S}_{10}$, and $\mathcal{S}_{11}$ as the LO term, $\mathcal{O}(\alpha_s)$ correction, $\mathcal{O}(v^2)$ correction, and $\mathcal{O}(\alpha_sv^2)$ correction, respectively, and we refer to the complete results including all corrections as the NLO results.
Note that although the right side of Eq. (\ref{eq_sij}) explicitly contains a $1/\epsilon_\text{IR}$ term, it cancels against the residue singularities from $\mathcal{M}_{11}$, resulting in a finite $\mathcal{S}_{11}$.

We now clarify the choices of the quarkonium mass in the calculation.
For the mass $M$ in the prefactor of Eq. (\ref{eq_amp1}), we use the relation $M=2\sqrt{m^2+\langle\boldsymbol{q}^2\rangle}$ \cite{Gremm:1997dq}, and expand it in powers of $\langle\boldsymbol{q}^2\rangle/m^2$.
This treatment enables a partial cancellation between the $\sqrt{2M}$ in Eq. (\ref{eq_amp1}) and the $2E_{\boldsymbol{q}}$ in Eq. (\ref{eq_pnr1}), thereby reducing the theoretical uncertainty from the charm quark mass $m$ \cite{Bodwin:2007ga}.
For the phase space computation in Eq. (\ref{eq_crss}), we instead use the physical $J/\psi$ mass.
This choice minimize the theoretical uncertainty further, as the $J/\psi$ mass are known very precisely.
Note, to maintain consistency, when determine the input parameters, the physical $J/\psi$ mass $M$, the charm quark mass $m$, and the matrix elements ratio $\langle\boldsymbol{q}^2\rangle$ are not treated as independent, they are constrained by the relation $M=2\sqrt{m^2+\langle\boldsymbol{q}^2\rangle}$.

\section{Details in the perturbative calculation}
\begin{figure}
\centering
\includegraphics[width=0.8\textwidth]{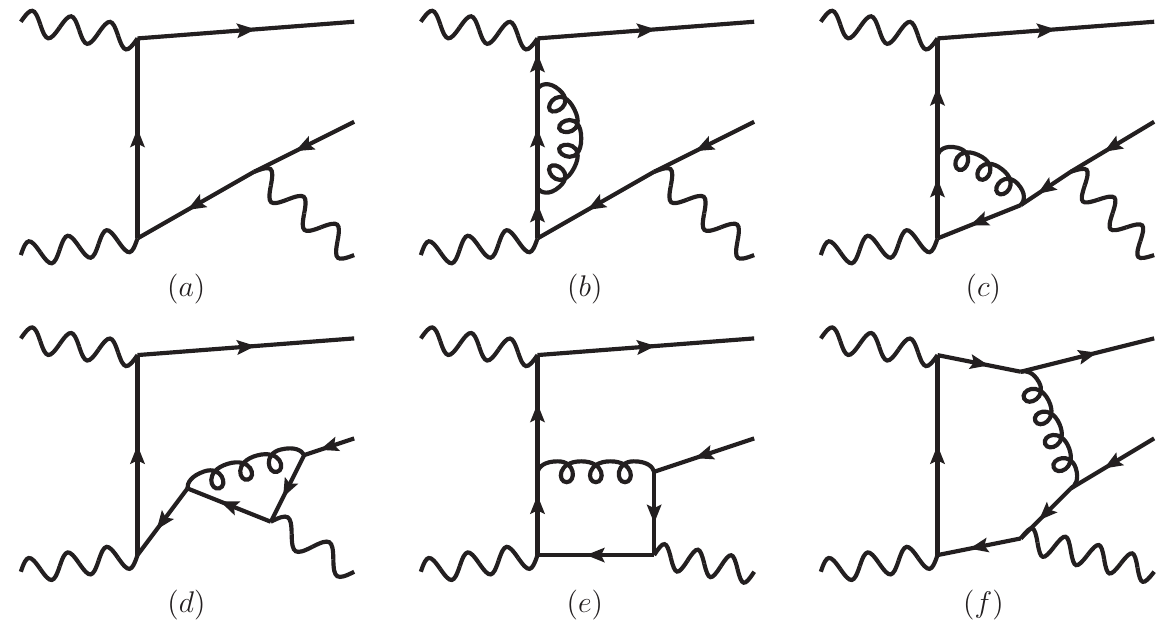}
\caption{Typical Feynman diagrams for $\gamma+\gamma\to J/\psi+\gamma$ process. (a) Tree-level diagram; (b)-(f) one-loop diagrams. Other diagrams can be obtained by exchanging the photon lines or inverting the quark line.}
\label{fig_feyn}
\end{figure}

In this section, we elucidate some key points in the calculation of $\mathcal{M}^{\gamma+\gamma\to c\bar{c}(^3S_1^{[1]})+\gamma}\big|_\text{pert QCD}$.
Considering the partonic process
\begin{equation}
\gamma(p_1)+\gamma(p_2)\to c\left(\frac{k}{2}+q\right)\bar{c}\left(\frac{k}{2}-q\right)+\gamma(p_3),
\end{equation}
where the momenta of external particles are indicated in parentheses.
The on-shell condition requires that $p_1^2=p_2^2=p_3^2=0$, $(\frac{k}{2}+q)^2=(\frac{k}{2}-q)^2=m^2$.
The typical Feynman diagrams involved in the calculation are shown in Fig. \ref{fig_feyn}.

To proceed the calculation, we work in the $c\bar{c}$ rest frame.
Choosing the direction of $\boldsymbol{p}_3$ as the $z$-axis, the explicit expressions of the external momenta are given by
\begin{align}
\label{eq_mm1}
&k=2E_{\boldsymbol{q}}(1,0,0,0),\quad q=(0,q_x,q_y,q_z),\\
\label{eq_mm2}
&p_3=\frac{2E_{\boldsymbol{q}} \sin(\theta_1+\theta_2)}{\sin\theta_1+\sin\theta_2-\sin(\theta_1+\theta_2)}(1,0,0,1),\\
\label{eq_mm3}
&p_1=\frac{2E_{\boldsymbol{q}} \sin\theta_2}{\sin\theta_1+\sin\theta_2-\sin(\theta_1+\theta_2)}(1,\sin\theta_1,0,\cos\theta_1),\\
\label{eq_mm4}
&p_2=\frac{2E_{\boldsymbol{q}} \sin\theta_1}{\sin\theta_1+\sin\theta_2-\sin(\theta_1+\theta_2)}(1,-\sin\theta_2,0,\cos\theta_2),
\end{align}
where $\theta_1$ and $\theta_2$ denote polar angles of $\boldsymbol{p}_1$ and $\boldsymbol{p}_2$, respectively.
For latter use, we introduce the rescaled momenta $\tilde{p}_1=p_1/(2E_{\boldsymbol{q}})$, $\tilde{p}_2=p_2/(2E_{\boldsymbol{q}})$, $\tilde{p}_3=p_3/(2E_{\boldsymbol{q}})$, $\tilde{k}=k/(2E_{\boldsymbol{q}})$, $\tilde{q}=q/\sqrt{\boldsymbol{q}^2}$, and the Mandelstam variables $\tilde{t}=(\tilde{p}_2-\tilde{p}_3)^2$, $\tilde{u}=(\tilde{p}_1-\tilde{p}_3)^2$.

We employ the covariant projection method to project out the spin-triplet, color-singlet amplitude.
The standard projector is \cite{Bodwin:2002cfe}:
\begin{equation}
\Pi_3 =\frac{1}{4\sqrt{2}E_{\boldsymbol{q}}(E_{\boldsymbol{q}}+m)}\left(\frac{\slashed{k}}{2}-\slashed{q}-m\right)\slashed{\epsilon}^{*}\left(\slashed{k}+2E_{\boldsymbol{q}}\right)\left(\frac{\slashed{k}}{2}-\slashed{q}+m\right)\otimes(\frac{\bf{1}_{c}}{\sqrt{N_{c}}}),
\end{equation}
where $\epsilon$ is the polarization vector of the spin-triplet state, and $\bf{1}_{c}$ denotes the unit color matrix.
When expanding the amplitude in powers of $\boldsymbol{q}^2$, note that the rescaled momenta ($\tilde{p}_1$, $\tilde{p}_2$, $\tilde{p}_3$, $\tilde{k}$, $\tilde{q}$), and polarization vectors (absent in Eqs. (\ref{eq_mm1})-(\ref{eq_mm4})), are $\boldsymbol{q}^2$-independent.
This renders the $\boldsymbol{q}^2$ expansion straightforward.
The amplitude is subsequently averaged over the spatial direction of $\boldsymbol{q}$, by utilizing the relations
\begin{align}
\int \frac{\text{d} \Omega}{4\pi}\;\tilde{q}^\mu=0,\quad \int \frac{\text{d} \Omega}{4\pi}\;\tilde{q}^\mu\tilde{q}^\nu=\frac{-g^{\mu\nu}+\tilde{k}^\mu\tilde{k}^\nu}{D-1},
\end{align}
where $D$ is the space-time dimensions.

At the LO in $\alpha_s$, there are six Feynman diagrams, one of which is shown in Fig. (\ref{fig_feyn})(a).
The calculation at this order is straightforward, yielding relatively compact results:
\begin{align}
\overline{\sum}\left|\mathcal{M}_{00}\right|^2=&\frac{131072 \pi^3 \alpha ^3 }{243 m^2 (\tilde{t}-1)^2 (\tilde{u}-1)^2 (\tilde{t}+\tilde{u})^2}(\tilde{t}^4+2 \tilde{t}^3 \tilde{u}-2 \tilde{t}^3+3 \tilde{t}^2 \tilde{u}^2\nonumber \\
&-3\tilde{t}^2\tilde{u}+\tilde{t}^2+2\tilde{t}\tilde{u}^3-3\tilde{t}\tilde{u}^2+\tilde{t}\tilde{u}+\tilde{u}^4-2 \tilde{u}^3+\tilde{u}^2),\\
\overline{\sum}2\text{Re}\left(\mathcal{M}_{10}\mathcal{M}_{00}^*\right)=&-\frac{131072 \pi^3 \alpha^3 }{729 m^2 (\tilde{t}-1)^3 (\tilde{u}-1)^3 (\tilde{t}+\tilde{u})^3}(4 \tilde{t}^6 \tilde{u}-2 \tilde{t}^6+12 \tilde{t}^5 \tilde{u}^2-18 \tilde{t}^5 \tilde{u}\nonumber \\
&+6 \tilde{t}^5+20 \tilde{t}^4 \tilde{u}^3-37 \tilde{t}^4 \tilde{u}^2+30 \tilde{t}^4 \tilde{u}-3 \tilde{t}^4+20 \tilde{t}^3 \tilde{u}^4-44 \tilde{t}^3 \tilde{u}^3+44 \tilde{t}^3 \tilde{u}^2\nonumber\\
&-16 \tilde{t}^3 \tilde{u}-4 \tilde{t}^3+12 \tilde{t}^2 \tilde{u}^5-37 \tilde{t}^2 \tilde{u}^4+44 \tilde{t}^2 \tilde{u}^3-16 \tilde{t}^2 \tilde{u}^2-2 \tilde{t}^2 \tilde{u}+3 \tilde{t}^2\nonumber \\
&+4 \tilde{t} \tilde{u}^6-18 \tilde{t} \tilde{u}^5+30 \tilde{t} \tilde{u}^4-16 \tilde{t} \tilde{u}^3-2 \tilde{t} \tilde{u}^2+2 \tilde{t} \tilde{u}-2 \tilde{u}^6+6 \tilde{u}^5-3 \tilde{u}^4\nonumber \\
&-4 \tilde{u}^3+3 \tilde{u}^2).
\end{align}
Here, $\tilde{t}$ and $\tilde{u}$ are Mandelstam variables defined previously.
Note that we use the physical $J/\psi$ mass in the phase-space calculation, while expanding the $c\bar{c}$ invariant mass $2E_{\boldsymbol{q}}$ in powers of $\boldsymbol{q}^2$ when calculating the squared amplitude.
This approach in fact introduces a projection between the momenta from the phase-space sampling, and those entering the squared amplitude calculation.
Our projection preserves the spatial directions of all external momenta in the $c\bar{c}$ (or $J/\psi$) rest frame.
Therefore, when performing the phase-space integral, the Mandelstam variables should be evaluated using $\tilde{t}=(\hat{p}_2-\hat{p}_3)^2/M^2$, $\tilde{u}=(\hat{p}_1-\hat{p}_3)^2/M^2$, where $\hat{p}_i$ denotes the momentum from the phase-space sampling.

At the NLO in $\alpha_s$, there are forty-eight one-loop diagrams, some of which are shown in Fig. (\ref{fig_feyn})(b)-(f).
We encounter both ultraviolet (UV) and infrared (IR) singularities.
The conventional dimensional regularization with $D=4-2\epsilon$ is employed to regularize them.
The UV singularities, which are contained in self-energy and triangle diagrams, are  removed by the renormalization procedure.
The relevant renormalization constants include $Z_2$ and $Z_m$, corresponding to heavy quark field and heavy quark mass respectively. 
We define them in the on shell (OS) scheme, with the counterterms given by
\begin{align}
\delta Z_2^{\rm OS}=&-C_F\frac{\alpha_s}{4\pi}\left[\frac{1}{\epsilon_{\rm UV}}+\frac{2}{\epsilon_{\rm IR}}-3\gamma_E+3\ln\frac{4\pi\mu_\text{R}^2}{m^2}+4\right], \\
 \delta Z_m^{\rm OS}=&-C_F\frac{\alpha_s}{4\pi}\left[\frac{3}{\epsilon_{\rm UV}}-3\gamma_E+3\ln\frac{4\pi\mu_\text{R}^2}{m^2} +4\right].
\end{align}
For the IR singularities, they completely cancel at $\mathcal{O}(\alpha_s v^0)$, while partially cancel at $\mathcal{O}(\alpha_s v^2)$.
The residual IR divergences match the singular term in the NRQCD calculation (Eq. (\ref{eq_pnr1})).
Consequently, our final result---the quantity $\mathcal{S}_{11}$ defined in Eq. (\ref{eq_sij})---is finite, as expected.

As a cross-check to our calculation, we successfully reproduce Eq. (1) of Ref. \cite{Feng:2012by}, the $J/\psi\to 3\gamma$  decay width up to $\mathcal{O}(\alpha_s v^2)$.

\section{Numerical results}
\subsection{Input parameters}

\begin{table}
   \centering
  \caption{Summary of parameters for photon-photon collisions in ultraperipheral Pb-Pb and p-p collisions at the HL-LHC. For each colliding system, we list the nucleon-nucleon center-of-mass energy $\sqrt{s_\text{NN}}$, beam energy $E_\text{beam}$, photon ``maximum'' energy $E_\gamma^\text{max}$, and effective charge radius $R$.}
  \label{tab_para}
    \begin{tabular}{p{2cm}<{\centering} p{2cm}<{\centering} p{2cm}<{\centering}  p{2cm}<{\centering} p{1.5cm}<{\centering}}
    \hline
   System & $\sqrt{s_\text{NN}}$ & $E_\text{beam}$ &$E_\gamma^\text{max}$ & $R$ \cr \hline
   Pb-Pb & $5.52$ TeV & $2.76$ TeV & $80$ GeV & 7.1 fm \cr
   p-p & $14$ TeV & $7.0$ TeV & $2.45$ TeV & 0.7 fm \cr \hline
    \end{tabular}
\end{table}

We investigate the UPC photon fusion process $A+B\to A+B+ J/\psi+\gamma$ at the HL-LHC.
Within the equivalent photon approximation framework \cite{vonWeizsacker:1934nji,Williams:1934ad}, the final cross section is expressed as a convolution of the $\gamma+\gamma\to J/\psi+\gamma$ subprocess cross section with the equivalent photon spectral functions:
\begin{align}
	\sigma^{A+B\to A+B+J/\psi+\gamma}= \int \frac{\text{d}x_1}{x_1} \frac{\text{d}x_2}{x_2}n_1(x_1)n_2(x_2) \hat{\sigma}^{\gamma+\gamma\to J/\psi+\gamma},
	\label{eq_fm}
\end{align}
where $x_i=E_{\gamma,i}/E_\text{beam}$ is the energy fraction of each photon, and $n_i(x)$ is the photon spectral function, for which we utilize the formula \cite{Cahn:1990jk,Shao:2022cly}
\begin{align}
n(x) = \dfrac{2 Z^2 \alpha}{\pi}\left\{\chi K_0(\chi)K_1(\chi) -(1-\gamma_\text{L}^{-2}) \dfrac{\chi^2}{2}\left[K_1^2(\chi)-K_0^2(\chi)\right]\right\}.
\end{align}
Here, $Z$ is the charge number of the ion, $K_i$ is the modified Bessel function, and $\gamma_\text{L}=E_\text{beam}/m_\text{p}$ is the Lorentz boost factor, with $m_\text{p}$ denoting the proton mass.
The variable $x$ enters through $\chi=xm_\text{p}R$, where $R$ is the effective charge radius of the nucleus.
Our study focuses on Pb-Pb and p-p collision modes; the relevant parameters \cite{Shao:2022cly,Bruce:2018yzs} are  summarized in Table \ref{tab_para}.

Due to the finite coverage of detectors, proper kinematic cuts should be imposed on the final state particles. 
Guided by the CMS measurement of $\gamma\gamma\to \gamma\gamma$ scattering \cite{CMS:2024bnt} and prior theoretical work \cite{Goncalves:2023sts}, we impose transverse momentum and rapidity cuts of $p_\text{T}>2\;\text{GeV}$ and $|y_{\psi,\gamma}|<2.4$ on both the final-state $J/\psi$ and $\gamma$ particles.
Importantly, the transverse momentum cut serves to suppress the background from the Pomeron exchange process and meanwhile facilitate the exclusive event selection.

Other parameters used in our numerical evaluation are as follows:
\begin{align}
&\alpha=1/137.065,\quad M=3.096\; \text{GeV},\quad m_\text{p}=0.9315\; \text{GeV},\nonumber \\
& \langle\mathcal{O}_1 \rangle_{J/\psi}=0.440^{+0.067}_{-0.055}\; \text{GeV}^3,\quad \langle \boldsymbol{q}^2\rangle=0.441^{+0.140}_{-0.140}\; \text{GeV}^2.
\end{align}
Here, the LDME and the ratio of matrix elements are taken from Ref. \cite{Bodwin:2007fz}, while other parameters are from the Particle Data Group's Review of Particle Physics \cite{ParticleDataGroup:2024cfk}.
In our calculation, the charm quark mass $m$ is not an independent input, but determined through the relation $M=\sqrt{m^2+\langle \boldsymbol{q}^2\rangle}$.
This yields $m=1.40^{+0.05}_{-0.05}$ GeV, where the center value agrees with the value commonly used in the literature.

The renormalization scale $\mu_\text{R}$ enters only through the strong coupling $\alpha_s(\mu_\text{R}^2)$.
We set $\mu_\text{R}=r_1\sqrt{4m^2+p_\text{T}^2}$, taking $r_1=1$ as the central value.
To estimate the renormalization scale uncertainty, we vary $r_1$ within the range $1/2<r_1<2$.
The factorization scale $\mu_\text{F}$, however, enters only at $\mathcal{O}(\alpha_s v^2)$, via the $\ln (\mu_\text{F}/m)$ term appearing in $\mathcal{S}_{11}$.
Therefore, we set $\mu_\text{F}=r_2m$, also with $r_2=1$ as the central value, and vary $r_2$ between $1/2$ and $2$ to estimate the factorization scale uncertainty.

For the running of the strong coupling constant, we adopt the two-loop formula \cite{Deur:2016tte}
\begin{equation}
\frac{4\pi}{\alpha_s(\mu_\text{R}^2)}-\frac{\beta_1}{\beta_0}\ln\left(\frac{4\pi}{\alpha_s(\mu_\text{R}^2)}+\frac{\beta_1}{\beta_0}\right) =\frac{4\pi}{\alpha_s(\mu_0^2)}-\frac{\beta_1}{\beta_0}\ln\left(\frac{4\pi}{\alpha_s(\mu_0^2)} +\frac{\beta_1}{\beta_0}\right)+\beta_0\ln\frac{\mu_\text{R}^2}{\mu_0^2},
\end{equation}
where $\beta_0=(11/3)C_A-(4/3)T_Fn_f$, $\beta_1=(34/3)C_A^2-4C_FT_Fn_f-(20/3)C_AT_Fn_f$ , with $C_A=3$, $C_F=4/3$, $T_F=1/2$.
We take $n_f=4$ active flavors, and set the initial scale $\mu_0=m_\tau=1.777$ GeV, with $\alpha_s(\mu_0^2)=0.314$ \cite{ParticleDataGroup:2024cfk}.

\subsection{Cross sections}
In Table \ref{tab_intCro}, we present the cross sections for the exclusive production of $J/\psi+\gamma$ via photon-photon fusion in Pb-Pb and p-p UPCs.
The column labeling ``Central'' contains the cross sections for central values of parameters.
One readily observes that the contributions of $\mathcal{O}(\alpha_s)$, $\mathcal{O}(v^2)$, and $\mathcal{O}(\alpha_s v^2)$ terms are about $-50\%$, $-33\%$, and $15\%$ of the LO term, which suggests reasonable convergence behavior in the NRQCD expansion for both $\alpha_s$ and $v^2$.
This pattern is in contrast to the situation of $J/\psi\to 3\gamma$, where the $\mathcal{O}(\alpha_s)$ and $\mathcal{O}(\alpha_s v^2)$ corrections are strikingly significant.
In all, these correction terms suppress the LO cross sections by a factor of about $1/3$, which is essential for making reliable phenomenological predictions.

\begin{table}
   \centering
  \caption{The total cross sections for the exclusive production of $J/\psi+\gamma$ via photon-photon fusion in Pb-Pb and p-p UPCs.The column labeling ``Central'' contains the cross sections for central values of parameters. Subsequent columns contain the theoretical uncertainties caused by the variation of corresponding parameters. The kinematic cuts $p_\text{T}>2\;\text{GeV}$ and $|y_{\psi,\gamma}|<2.4$ are imposed in all cases.}
  \label{tab_intCro}
    \begin{tabular}{|p{2cm}<{\centering}|p{2.5cm}<{\centering}| p{1.5cm}<{\centering} p{1.5cm}<{\centering} p{1.5cm}<{\centering} p{1.5cm}<{\centering} p{1.5cm}<{\centering}|}
    \hline
   System & Term &Central & $\Delta \langle \boldsymbol{q}^2\rangle$ & $\Delta \langle \mathcal{O}_1 \rangle$ & $\Delta \mu_\text{R}$ & $\Delta \mu_\text{F}$ \cr\hline
  \multirow{5}{*}{\makecell[c]{Pb-Pb, \\ $\sqrt{s_\text {NN}}=$\\5.52 TeV}}& LO (nb) & $87.90$ & $^{+10.38}_{-8.62}$ & $^{+13.38}_{-10.99}$ & --- & ---\cr\cline{2-7}
 &  $ \mathcal{O}(\alpha_s)$ (nb)& $-43.93$ & $^{-5.49}_{+4.59}$ & $^{-6.69}_{+5.49}$ & $^{+8.29}_{-13.72}$ & ---\cr\cline{2-7}
&  $ \mathcal{O}(v^2)$ (nb) &  $-29.08$ &$^{-17.05}_{+12.38}$ & $^{-4.43}_{+3.64}$ & --- & ---\cr\cline{2-7}
& $ \mathcal{O}(\alpha_s v^2)$ (nb)& $12.98$  & $^{+7.74}_{-5.57}$ & $^{+1.98}_{-1.62}$ & $^{-2.44}_{+4.04}$ & $^{+1.78}_{-1.79}$\cr\cline{2-7}
& NLO (nb) & $27.88$ & $^{-4.42}_{+2.77}$ & $^{+4.24}_{-3.48}$ & $^{+5.85}_{-9.68}$& $^{+1.78}_{-1.78}$\cr \hline\hline
  \multirow{5}{*}{\makecell[c]{p-p, \\ $\sqrt{s_\text {NN}}=$\\14 TeV}}&LO (fb) & $10.970$ & $^{+1.292}_{-1.080}$ & $^{+1.670}_{-1.371}$ & --- & --- \cr\cline{2-7}
 &  $ \mathcal{O}(\alpha_s)$ (fb) & $-5.462$ & $^{-0.683}_{+0.569}$ & $^{-0.832}_{+0.683}$ & $^{+1.025}_{-1.698}$ & ---\cr\cline{2-7}
&  $ \mathcal{O}(v^2)$ (fb)&  $-3.646$ & $^{-2.136}_{+1.552}$ & $^{-0.555}_{+0.456}$ & --- & ---\cr\cline{2-7}
& $ \mathcal{O}(\alpha_sv^2)$ (fb) & $1.627$  & $^{+0.970}_{-0.698}$ & $^{+0.248}_{-0.203}$ & $^{-0.304}_{+0.503}$ & $^{+0.222}_{-0.222}$\cr\cline{2-7}
&NLO (fb) & $3.489$ & $^{-0.557}_{+0.342}$ & $^{+0.531}_{-0.436}$ & $^{+0.720}_{-1.194}$& $^{+0.222}_{-0.222}$\cr \hline
    \end{tabular}
\end{table}

In addition to the central values, the theoretical uncertainties that are associated with the variations of the ratio of the matrix elements $\langle \boldsymbol{q}^2\rangle$, the LDME $\langle \mathcal{O}_1\rangle$, the renormalization scale $\mu_\text{R}$, and the factorization scale $\mu_\text{F}$ are separately listed in Table \ref{tab_intCro}.
Since the $J/\psi$ mass is known very precisely, we omit its variation from our uncertainty analysis. 
Adding all uncertainties in quadrature, we obtain
\begin{align}
&\sigma_{\rm LO}=87.90^{+16.94}_{-13.97}\;\text{nb},\quad \sigma_{\rm NLO}=27.88^{+7.94}_{-11.34}\;\text{nb},\quad \text{for Pb-Pb collision};\\
&\sigma_{\rm LO}=10.970^{+2.112}_{-1.746}\;\text{fb},\quad \sigma_{\rm NLO}=3.489^{+0.984}_{-1.406}\;\text{fb},\quad \text{for p-p collision}.
\end{align}
The results show that the theoretical uncertainties increase after including the QCD and relativistic corrections, which can be explained by the fact that the renormalization and factorization scales are absent in the LO calculation.

The HL-LHC will substantially increase the amount of ion collisions delivered to the LHC experiments.
In a typical operation year, the integrated luminosities are about $5\;\text{nb}^{-1}$ for Pb-Pb collision \cite{Bruce:2018yzs} \footnote{This assumes one month per year is allocated to Pb-Pb collision.}, and $150\;\text{fb}^{-1}$ for p-p collision \cite{Shao:2022cly}.
Assuming $J/\psi$ meson is reconstructed through the decay $J/\psi\to \ell^+\ell^-\;(\ell=e,\; \mu)$ with a branching fraction of about $12\%$ \cite{ParticleDataGroup:2024cfk}, and a photon reconstruction and identification efficiency of about $40\%$ \cite{CMS:2024bnt}, the number of reconstructed $J/\psi+\gamma$ candidates per operation year is about 4-8 for Pb-Pb collision, and about 15-32 for p-p collision.
Although the number of candidates is moderate, the clear event topology makes the experimental investigations feasible.
Moreover, as we will show latter, the number of events could increase substantially by lowering the minimum transverse momentum cut.

\begin{figure}
\centering
\begin{subfigure}{0.45\textwidth}
\includegraphics[width=\textwidth]{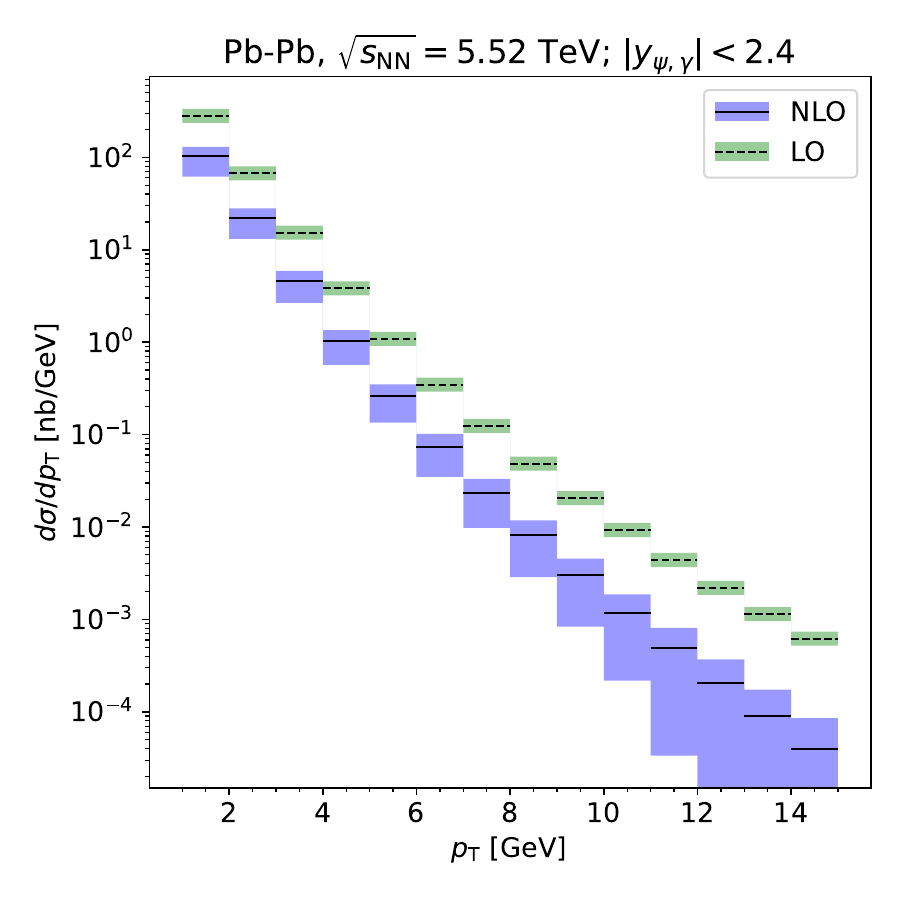}
\caption{}
\end{subfigure}
\begin{subfigure}{0.45\textwidth}
\includegraphics[width=\textwidth]{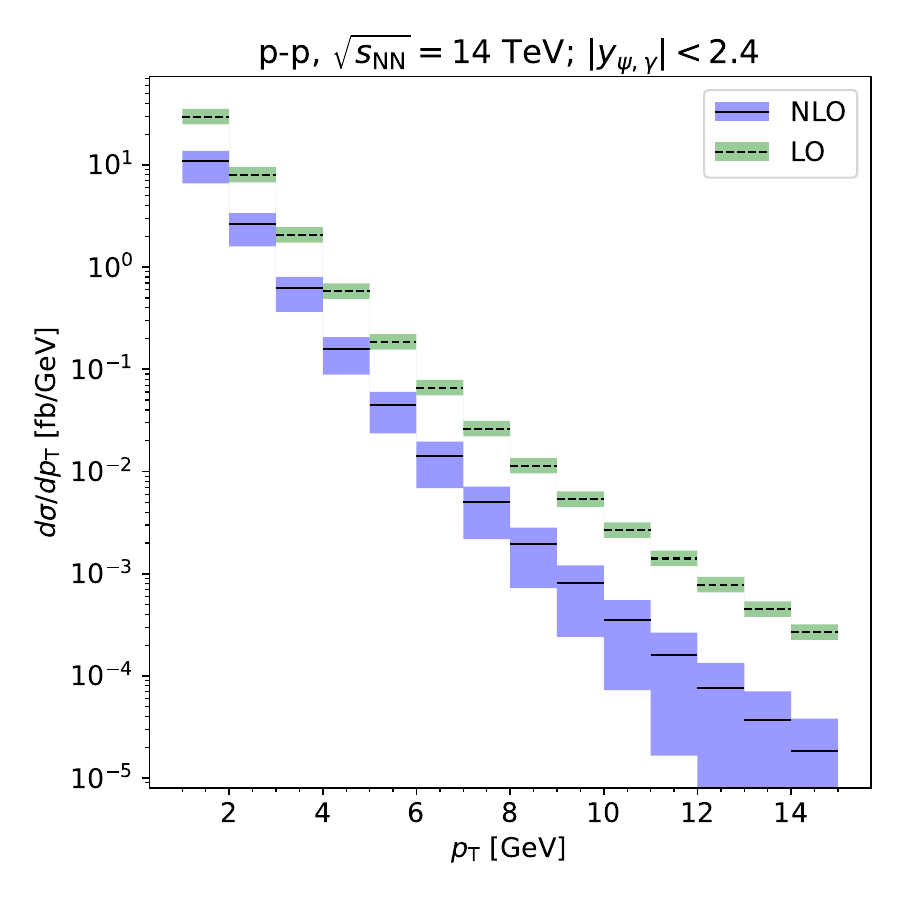}
\caption{}
\end{subfigure}
\caption{Differential cross sections with respect to the $J/\psi$ transverse momentum $p_\text{T}$ for (a) Pb-Pb collision and  (b) p-p collision. Theoretical uncertainties are shown as color bands.}
\label{fig_1dis}
\end{figure}

Differential cross sections with respect to the $J/\psi$ transverse momentum $p_\text{T}$ for Pb-Pb and p-p collisions are shown in Fig. \ref{fig_1dis}(a) and (b), respectively.
Both the LO and NLO distributions decrease steeply and monotonically as $p_\text{T}$ increases from 1 to 15 GeV.
Recall that our integrated cross-section calculations employ a transverse momentum cut of $p_\text{T}> 2$ GeV.
As Fig. \ref{fig_1dis} shows, lowering this cut to $p_\text{T}> 1$ GeV would quadruple the integrated cross sections.
Another thing to note is that when setting the renormalization scale to $\mu_\text{R}=\sqrt{4m^2+p_\text{T}^2}/2$, the NLO differential cross sections become negative in the last three bins.
However, since the contributions from this region are not dominant, and the negative cross section issue does not appear in the central value calculation (where $\mu_\text{R}=\sqrt{4m^2+p_\text{T}^2}$), no additional treatment is applied to address this problems.

\begin{figure}
\centering
\begin{subfigure}{0.45\textwidth}
\includegraphics[width=\textwidth]{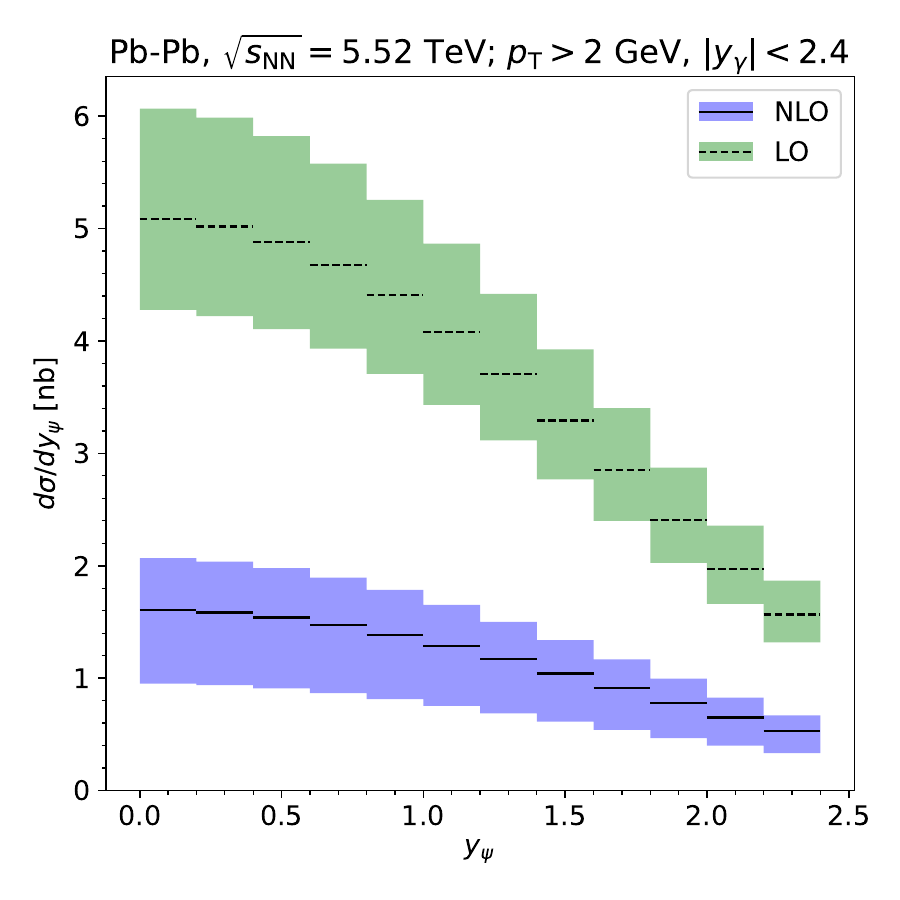}
\caption{}
\end{subfigure}
\begin{subfigure}{0.45\textwidth}
\includegraphics[width=\textwidth]{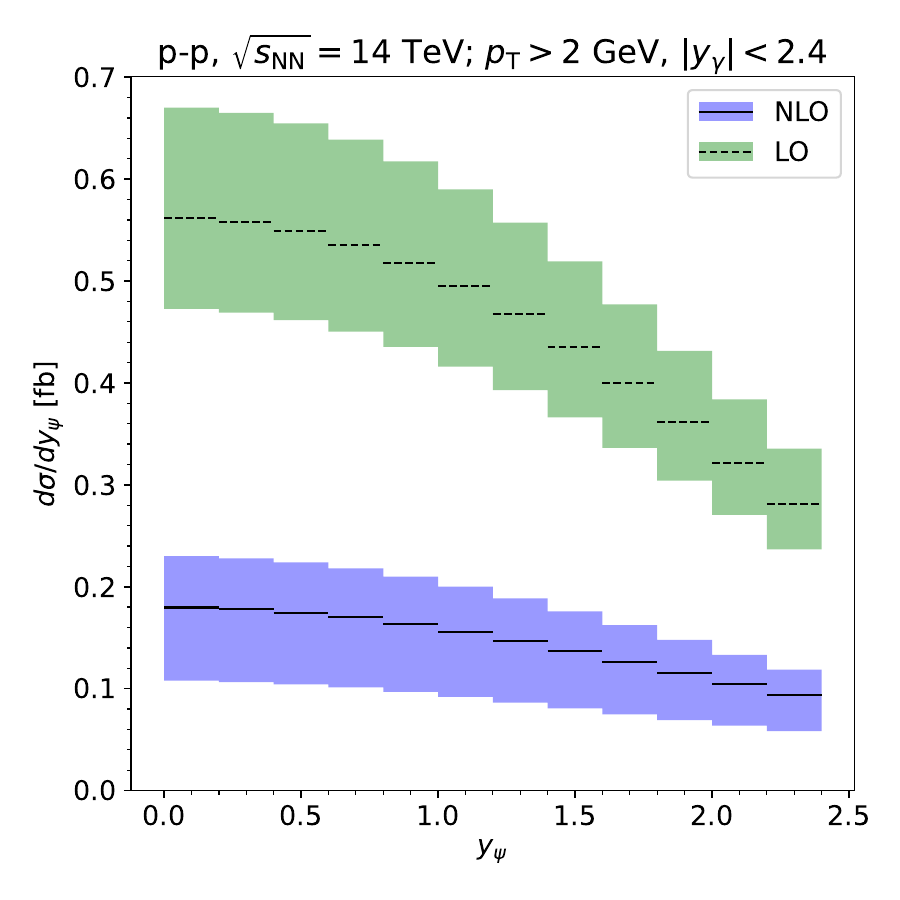}
\caption{}
\end{subfigure}
\caption{Differential cross sections with respect to the $J/\psi$ rapidity $y_\psi$ for (a) Pb-Pb collision and  (b) p-p collision. Theoretical uncertainties are shown as color bands.}
\label{fig_2dis}
\end{figure}

Differential cross sections with respect to the $J/\psi$ rapidity $y_\psi$ for Pb-Pb and p-p collisions are shown in Fig. \ref{fig_2dis}(a) and (b), respectively.
Since the distributions is symmetric over $y_\psi=0$, only the $y_\psi>0$ region is plotted.
 It can be seen that as $y_\psi$ increases from 0 to 2.4, both LO and NLO differential cross sections drop steadily.

\section{Summary}
In this work, we investigate the exclusive production of $J/\psi+\gamma$ in ultraperipheral Pb-Pb and p-p collisions at the HL-LHC.
Working within the NRQCD factorization framework, we calculate the $\mathcal{O}(\alpha_s)$, $\mathcal{O}(v^2)$, and  $\mathcal{O}(\alpha_sv^2)$ corrections to the exclusive process $\gamma+\gamma\to J/\psi+\gamma$.
We evaluate both total cross sections and differential distributions with respect to the $J/\psi$ transverse momentum and rapidity.
Numerical results show that the $\mathcal{O}(\alpha_s)$, $\mathcal{O}(v^2)$, and $\mathcal{O}(\alpha_s v^2)$ corrections represent approximately $-50\%$, $-33\%$, and $15\%$ of the LO contribution, exhibiting reasonable convergence behavior both in $\alpha_s$ and $v^2$ expansion.
Collectively, these corrections suppress the LO cross sections by a factor of about $1/3$, which is essential for reliable phenomenological predictions.
With appropriate kinematic cuts, our final results including all corrections are $\sigma_{\rm NLO}=25.18^{+7.18}_{-10.34}\;\text{nb}$ for Pb-Pb collision, and $\sigma_{\rm NLO}=3.053^{+0.865}_{-1.238}\;\text{fb}$ for p-p collision. 
These values suggest that a future experimental analysis is feasible.
Finally, we remark that further exploration of the $\mathcal{O}(\alpha_s^2)$ and $\mathcal{O}(v^4)$ corrections to this process may turn out to be enlightening.

\vspace{2.0cm} {\bf Acknowledgments}
This work was supported by the National Natural Science Foundation of China (NSFC) under the Grants 12205061 and 12175048. The work of LBC was also supported by the Guangdong Basic and Applied Basic Research Foundation  under grant No.  2025B1515020009.
\vspace{2.0cm}

\end{document}